\documentclass[conference]{IEEEtran}

\usepackage{xspace}

\usepackage{ifthen}
\providecommand\BLIND{FALSE}
\ifthenelse{\equal{\BLIND}{FALSE}}{
  \newcommand{\AUTHORBLOCK}{%
    \author{\IEEEauthorblockN{%
        Neal Master\IEEEauthorrefmark{1}\IEEEauthorrefmark{2}\IEEEauthorrefmark{4}, %
        David Scheinker\IEEEauthorrefmark{3}\IEEEauthorrefmark{4}, and %
        Nicholas Bambos\IEEEauthorrefmark{1}\IEEEauthorrefmark{3}\\
        \texttt{\{nmaster, dscheink, bambos\}@stanford.edu}}%
      \IEEEauthorblockA{\IEEEauthorrefmark{1}Department of Electrical Engineering, Stanford University, Stanford, CA 94305}%
      \IEEEauthorblockA{\IEEEauthorrefmark{2}Department of Statistics, Stanford University, Stanford, CA 94305}%
      \IEEEauthorblockA{\IEEEauthorrefmark{3}Department of Management Sciences \& Engineering, Stanford University, Stanford, CA 94305}%
      \IEEEauthorblockA{\IEEEauthorrefmark{4}Lucile Packard Children's Hospital Stanford, Palo Alto, CA 94304}
    }%
  }
  \newcommand{\ACKNOWLEDGEMENTS}{%
    \section{Acknowledgments}
    We would like to thank the nurses and physicians at LPCH for sharing
    their domain expertise. In particular, we acknowledge Beverly Schuler,
    Johannon Olson, and Craig Albanese. We would also like to thank the
    Chief Operating Officer of LPCH, Anne McCune, for encouraging this
    research.
    
    Neal Master is funded by a Stanford Graduate Fellowship (SGF) in
    Science \& Engineering.%
  }
  \newcommand{\LPCHFULL}{Lucile Packard Children's Hospital (LPCH) Stanford\xspace}
  \newcommand{\LPCH}{LPCH\xspace}
}{
  \newcommand{\AUTHORBLOCK}{\author{\IEEEauthorblockN{\color{red}{\it DRAFT ANONYMIZED FOR DOUBLE-BLIND REVIEW}}}}
  \newcommand{\ACKNOWLEDGEMENTS}{}
  \newcommand{\LPCHFULL}{our pediatric hospital\xspace}
  \newcommand{\LPCH}{our pediatric hospital\xspace}
}

\usepackage{graphicx}
\usepackage{amsmath, amsfonts}
\usepackage[caption=false,font=footnotesize]{subfig}
\usepackage{url}

\usepackage{color}

\newcommand{\AVG}{\textbf{AVG}\xspace}
\newcommand{\SCH}{\textbf{SCH}\xspace}
\newcommand{\DTR}{\textbf{DTR}\xspace}
\newcommand{\RFR}{\textbf{RFR}\xspace}
\newcommand{\ABR}{\textbf{ABR}\xspace}
\newcommand{\DTRSCH}{\textbf{DTR-SCH}\xspace}
\newcommand{\RFRSCH}{\textbf{RFR-SCH}\xspace}
\newcommand{\ABRSCH}{\textbf{ABR-SCH}\xspace}

\newcommand{\abs}[1]{\left|#1\right|}
\newcommand{\eps}{\epsilon}
\newcommand{\set}[1]{\left\{#1\right\}}

\usepackage{booktabs}

\usepackage{array}
\newcolumntype{C}[1]{>{\centering\let\newline\\\arraybackslash\hspace{0pt}}m{#1}}
\newcolumntype{L}[1]{>{\raggedright\let\newline\\\arraybackslash\hspace{0pt}}m{#1}}

\usepackage{hhline}

\begin{document}

\title{Predicting Pediatric Surgical Durations}

\AUTHORBLOCK

\maketitle

\begin{abstract}
  Effective management of operating room resources relies on accurate
  predictions of surgical case durations. This prediction problem is
  known to be particularly difficult in pediatric hospitals due to the
  extreme variation in pediatric patient populations.  We propose a
  novel metric for measuring accuracy of predictions which captures
  key issues relevant to hospital operations. With this metric in mind
  we propose several tree-based prediction models. Some are automated
  (they do not require input from surgeons) while others are
  semi-automated (they do require input from surgeons). We see that
  many of our automated methods generally outperform currently used
  algorithms and even achieve the same performance as surgeons. Our
  semi-automated methods can outperform surgeons by a significant
  margin. We gain insights into the predictive value of different
  features and suggest avenues of future work.
\end{abstract}

\section{Introduction}

Operating rooms are a critical hospital resource that must be managed
effectively in order to deliver high quality of care at a reasonable
cost. Because of the expensive equipment and highly trained staff,
operating rooms are necessarily very expensive with the average cost
of operating room time in the US being roughly \$4000 per hour
\cite{Macario_2010, Shippert_2005}. In addition, mismanaged operating
rooms lead to cancelled surgeries with each cancellation decreasing
revenue by roughly \$1500 per hour \cite{Dexter_2005, Macario_2001,
  Dexter_2002}. The managment process is complicated and must account
for heterogeneity of patient needs, uncertainty in patient recovery
times, and uncertainty in surgical durations. In this paper, we aim to
design models that predict surgical durations with high accuracy. The
motivating idea is that better predictions enable better operational
decisions.

Specifically, we consider the problem of predicting pediatric surgical
durations. Currently, many pediatric hospitals rely on surgeons to
provide predictions and this alone increases costs. Not only is a
surgeon's individual time expensive, a surgeon may depend on the help
of their staff to make the predictions. Each of these contributions
may seem insignificant on its own, but these manhours add up to
increased costs. Consequently, although our primary goal is to
increase prediction accuracy, automating the prediction process even
without increasing accuracy can help reduce costs and improve
efficiency.

A major reason that pediatric hospitals rely on surgeons' medical
expertise is that accurately predicting pediatric surgical case
lengths is a very difficult problem. It is considered to be a more
difficult problem than predicting adult surgical durations because
compared to patient populations at adult hospitals, patient
populations at pediatric hospitals are characterized by extreme
variation in patient age, size, and developmental level. This has been
discussed in the academic medical literature for specific procedures
\cite{Smallman_2010} and is also supported by anecdotal evidence at
\LPCHFULL. Although we use data from \LPCH, our goal in is to design
models that apply to all pediatric hospitals. In particular, none of
the features used by our models are specific to \LPCH. Moreover, even
though we consider a multitude of different procedure types, we only
use features that are relevant to all kinds of surgical procedures. In
this sense, we aim to provide a ``one-size-fits-most'' solution that
is broadly applicable to pediatric hospitals regards of size or case
load profile.

Given this broad motivation, there are several papers on the topic of
predicting surgical duations. However, the majority are focused on
adult patient populations (e.g. \cite{Strum_Predict_2000,
  Stepaniak_2010, Wright_1996}) with pediatric populations only being
of very recent interest, e.g. \cite{Bravo_2015}. In addition, many of
these studies rely on simple methods like ordinary least squares
regression \cite{Strum_Predict_2000, Stepaniak_2010, Wright_1996};
regression trees are the most modern technique considered in the
literature \cite{Bravo_2015}. For adult surgeries, researchers
typically see ``modest improvements in accuracy'' \cite{Wright_1996}
over human experts (e.g. surgeons and nurses). In contrast, for
pediatric surgeries, the difficulty of the problem leads to negative
conclusions with \cite{Bravo_2015} reporting that ``none of the
variables previously associated with case time in adults were
generally correlated with case time in our pediatric
population''. These papers demonstrate that compared to predicting
adult surgical durations, predicting pediatric surgical durations is
still a difficult open problem.

The remainder of our paper is organized as follows. In
Section~\ref{sec:background} we discuss how surgical duration
predictions are made and used at \LPCH. We also discuss current
research on predictive models for pediatric surgical durations. In
Section~\ref{sec:metric} we motivate and define a performance metric
that we use to quantify prediction accuracy. This metric models
operational concerns in a hospital setting and motivates a nonlinear
transformation of the data. In Section~\ref{sec:results} we define
some benchmark prediction methods, propose our own prediction models,
and present a comparison. We find that our models outperform currently
used algorithms as well as expert predictions. This demonstrates that
(in spite of the results in \cite{Bravo_2015}), tree-based regression
methods can be used to predict pediatric surgical durations. In
Section~\ref{sec:future} we describe some directions of future
work. We conclude in Section~\ref{sec:conclusions}.

\section{The Importance of Accurate Predictions and the State of the
  Art\label{sec:background}}
In this section we provide background into the prediction problem at
hand. First we explain how predictions influence the scheduling of
surgical environments at hospitals like \LPCH. We then explain the
methods that are currently used to make these predictions. We discuss
the academic literature on the topic.

\subsection{The State of Practice\label{sec:current}}
Scheduling a surgical procedure can be a very complicated process for
the patient as well as for the hospital. Depending on the type of
procedure, the scheduling process may begin several weeks before the
procedure actually takes place. The patient and primary physician will
coordinate with the appropriate surgical service to meet the clinical
needs of the patient. Because patient preferences and hospital
policies play a big role in determining these coarse-grained
scheduling decisions, we will not describe them in detail. The most
important feature of the coarse-grained scheduling is that operating
rooms are shared across different surgical services with block
scheduling. This means that a particular surgical service will have an
operating room for a large block of time; at \LPCH a single block
constitutes the entire day.  Different block scheduling strategies can
be used, e.g. \cite{Zenteno_2015}, but regardless of how blocks are
allocated across the week or month, block scheduling causes many
similar procedures to be scheduled back-to-back.

The fine-grained scheduling decisions are made in the days immediately
preceding a surgery. The two-step process we describe is somewhat
specific to \LPCH but similar systems exist at other pediatric
surgical hospitals. The first step is the prediction: surgeons (with
varying levels of assistance from their administrative staff) will
need to predict the duration of each surgical procedure. The second
step is the scheduling: a group of nurses and physicians use the
predictions to schedule the operating rooms (ORs) and the Ambulatory
Procedure Unit (APU)\footnote{At \LPCH there are currently seven ORs
  and one APU.}.

Although the scheduling is done manually rather than by an
optimization algorithm, the nurses and physicians who make the
scheduling decisions have several objectives in mind. One is to have
all procedures completed by the end of the day; it is possible to run
overtime but this is inconvenient and incurs additional staffing
costs. Another objective is to have patients released to the recovery
units at a regular rate so that recovery beds are available and the
nursing staff is able to provide care.

The predictions impact all of these objectives and more. Consider a
scenario in which some surgeries finish earlier than predicted and
other surgeries finish later than predicted. Suddenly there is an
unexpected spike in demand for recovery beds that the nursing staff is
unable to accommodate. Patients will need to wait (in the ORs) for
beds and this delays other surgeries. If these delays are acute,
surgeries will need to be rescheduled. This operational inefficiency
reduces quality of care for patients and increases costs for the
hospital. Patients whose surgeries have been cancelled may opt to have
their procedures done at other hospitals. Thus, inaccurate predictions
not only increase costs but also reduce revenue.

The prediction process is essential to delivering high quality care
while maintaining efficiency, but the current prediction methods are
somewhat primitive. There are two competing methods available to most
pediatric surgical hospitals. The first is historical averaging. When
a particular surgeon is scheduled to perform a particular surgery,
historical averaging predicts that the surgical duration is the
average (arithmetic mean) of the past five times the surgeon performed
that procedure. This method does not take into account the substantial
variation in the patients and can be quite inaccurate. The second
method is to rely on expert opinions. Surgeons (potentially with the
assistance of their staff) can provide an estimate of how much time
they need to complete a given procedure. This is the system currently
used at \LPCH. Although surgeons have extensive experience, their
predictions are not necessarily accurate. One reason is that
physicians are not using quantitative models so their predictions are
merely ``guesstimates.'' Another reason is that physicians may have
financial incentives that cause them to systematically misestimate the
surgical durations. For example, surgeons do not bear the overtime
costs of surgeries running past the end of the day. As a result, in an
effort to maximize the number of surgeries scheduled in a block (and
hence maximize their personal compensation), surgeons may
underestimate the amount of time required to complete a procedure.

Given the inadequacies of both historical averaging and expert
predictions, it is not clear which method is superior. Not only does
this comparison depend on the types of procedures and the population
of patients, it also depends significantly on the surgical teams. At
\LPCH, expert prediction is currently used but this might not be the
best choice for other hospitals. As we develop and evaluate our
predictive models, we will need to compare our performance to both of
these existing benchmark practices.

\subsection{Literature Review\label{sec:review}}
Much of the applied statistics and academic medical literature on
surgical procedure durations focuses on modeling problems rather than
on prediction problems. For example, in \cite{Strum_2000} it was shown
that lognormal distributions model the uncertainty of surgical
procedure times better than normal distributions. A consequent line of
research explores different methods for fitting lognormal
distributions, e.g. \cite{May_2000, Spangler_2004}. Although this work
does not directly address the prediction problem and is not focused on
pediatrics, it does point out that surgical times tend to follow
heavy-tailed distributions. This insight is valuable when designing
predictive models and is discussed more in Section~\ref{sec:metric}.

The literature on predictive modeling for pediatric surgeries is
sparse; the primary paper on this topic is based on data from Boston
Children's Hospital \cite{Bravo_2015}. This pioneering work identifies
improving predictions of pediatric surgical durations\footnote{ We
  note that in \cite{Bravo_2015}, surgical durations are measured from
  ``wheels in'' to ``wheels out'', i.e. patient entry to the OR to
  patient exit from the OR. In our work, we focus on predicting the
  time from when the surgeon enters the OR to when the surgeon exits
  the OR.  Although slightly different, we note that the dominant
  source of variability is the surgery. At \LPCH, surgeons are asked
  to predict the amount of time they spend in the OR and this allows
  us to directly compare our methods to expert predictions.} as a key
avenue of research. However, the paper is pessimistic: the authors
find that ``for most procedure types, no useful predictive factors
were identified and, most notably, surgeon identity was unimportant.''
They use surgeon identity, intensive care unit bed request, ASA status
(explained in Section~\ref{sec:results}), patient age, and patient
weight as features and conclude that ``until better predictors can be
identified, scheduling inaccuracy will persist.''

The negative results in \cite{Bravo_2015} demonstrate that building
predictive models for pediatric surgical durations is very difficult
but we must raise some concerns with their statistical approach. Our
primary criticisms are that the authors rely on a single learning
algorithm and that they impose restrictions on this algorithm in a way
that inhibits it's performance. Specifically, the authors rely on the
CART algorithm \cite{CART}. However, the authors opt to learn a
separate tree for each procedure. This essentially forces the tree to
split until each node has only one procedure type and then CART is
used to learn the remainder of the tree. The motivating idea is that
the procedure name is a very important feature but this model
restriction unnecessarily fragments the data.  A related issue is that
for many procedures the authors had only 30 observations, creating
training sets of 20 observations and testing sets of 10
observations. Given these small sample sizes, the authors restrict the
learned trees to have depths of at most three. Although this model
restriction may be appropriate for procedures with small sample sizes,
for some procedures the authors had hundreds of observations and with
with larger sample sizes, deeper trees can be learned.

We also note that CART is inherently unstable, i.e. the learned tree
is very sensitive to the training data \cite{Breiman_1996}. As a
result, although it is easy to draw conclusions from the topology of a
learned decision tree, it is difficult to have confidence in these
conclusions. This is exacerbated by small sample sizes. Consequently,
the conclusions presented in \cite{Bravo_2015} should be viewed with
skepticism.

Despite our concerns with the statistical methodology of
\cite{Bravo_2015}, we think that a more fundamental issue is the
modeling methodology. The authors use mean square error (MSE) as the
information criterion for fitting their decision tree models. In
addition, they use root mean square error (RMSE), mean absolute error
(MAE), and mean absolute percentage error (MAPE) as performance
metrics. Although these metrics are statistically meaningful, they are
not necessarily operationally meaningful. Given that scheduling
decisions are made by human experts, we feel that performance metrics
should be more easily interpretable by physicians and nurses. We
address this concern in the following section.

\section{The Impact of a Novel Performance Metric\label{sec:metric}}
This study is motivated primarily by operational concerns so we focus
a performance metric that is operationally meaningful. Although RMSE
has a natural statistical meaning, it does not have an obvious
operational meaning. For example, it is unlikely that a physician
would ask for the emperical root-mean-square prediction error of a
proposed prediction model. However, it is likely that a physician
would ask how often the predictive model is ``correct''.  The nuance
now is quantifying what it means for a prediction to be
``correct''. With this in mind, in this section we propose a novel
performance metric that has been developed based on input from
physicians and nurses at \LPCH. Furthermore, we will see that this
performance metric motivates a particular nonlinear transformation of
the data.

To formulate an operationally meaningful performance metric, we need
to consider the inherent asymmetry of the prediction and the
outcome. The predictions are used to set daily schedules and surgeries
will be delayed or cancelled if the outcomes are sufficiently
dissimilar from the predictions. Since similar kinds of procedures
tend to be scheduled back-to-back within the same block, the threshold
at which schedules ``break down'' depends on how long the procedures
are predicted to take.  We demonstrate this with the following
hypotheticals. First suppose a particular procedure is predicted to
take 150 minutes. If the procedure actually takes 165 minutes, then it
is reasonable to say that the prediction was correct -- the procedure
only took 10\% longer than was predicted and this will not
significantly impact subsequent procedures which are scheduled for
comparable amounts of time. Now suppose a different procedure in a
different block is predicted to take 20 minutes. If the procedure
actually takes 35 minutes then it is not reasonable to claim that the
prediction was correct. Short procedures are typically scheduled
back-to-back and so if the procedure takes 75\% longer than predicted
then this will undoubtably cause operational problems. Note that in
both cases the outcome was 15 minutes longer than the prediction but
in one case the prediction was deemed correct and in the other case
the prediction is incorrect. This demonstrates that the difference
between the outcome and the prediction needs to be less than some
percentage of the prediction in order for the prediction to be deemed
accurate.

However, there are limits to this reasoning. Consider the same
hypotheticals as above. If we have a procedure that is predicted to
take 150 minutes, then having the outcome be within 10\% (15 minutes)
of this predicted time, is reasonable. However, if we consider a
procedure that is predicted to take 20 minutes then requiring that the
outcome be within 10\% (now just 2 minutes) is no longer
reasonable. Clearly, using a simple percentage is too restrictive for
surgeries that are typically short. Similarly, using a simple
percentage is too lax for surgeries that are typically quite long. 

\begin{figure}
  \centering
  \includegraphics[width=\columnwidth]{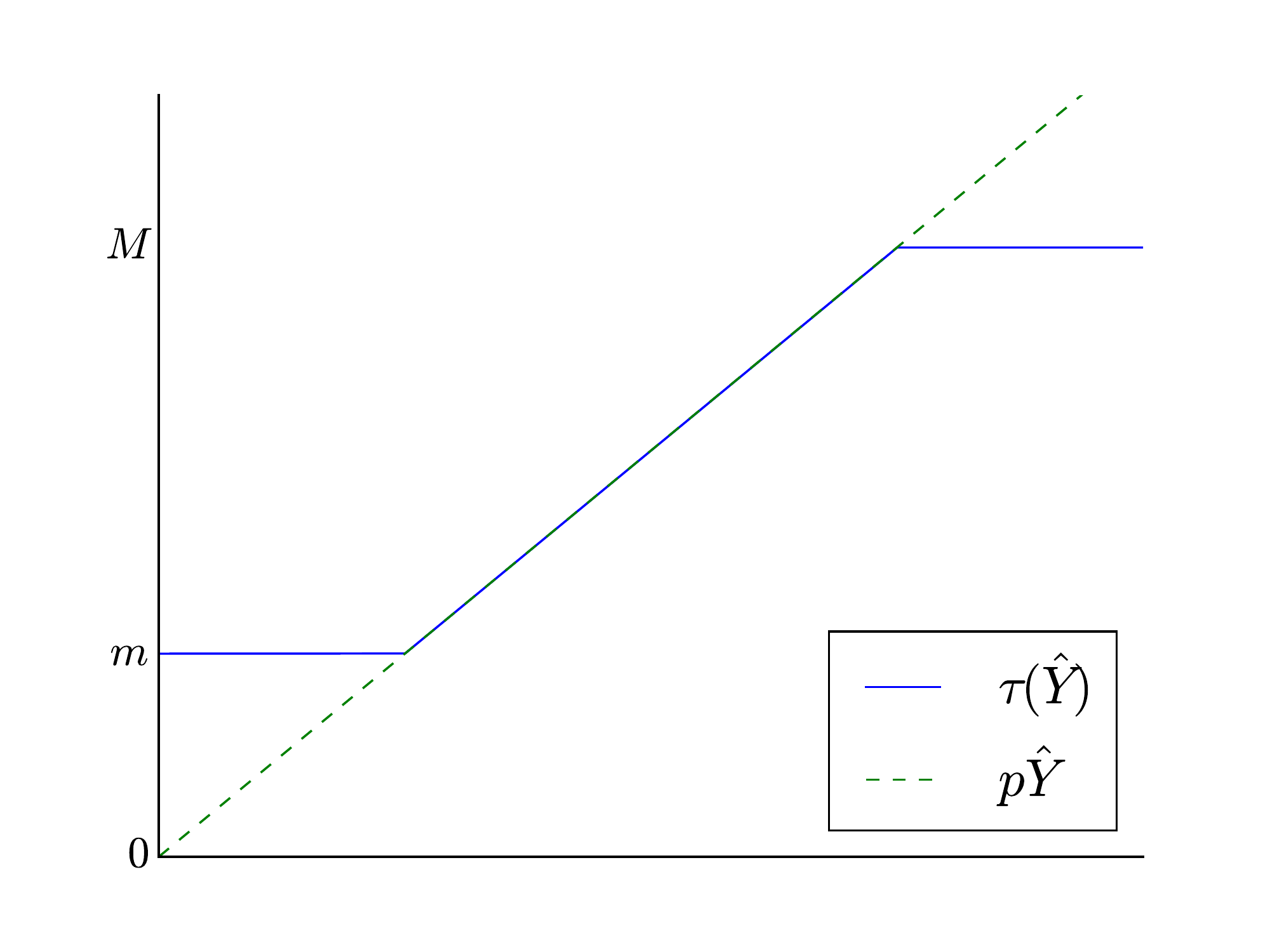}
  \caption{A depiction of the $\tau(\hat Y)$\label{fig:thresh}}
\end{figure}

To formalize the insights from these hypotheticals, we consider the
following model
\begin{equation}
  Y = f(X) + \eps
\end{equation}
where $X$ is a vector of features describing a particular surgical
case, $Y$ is the amount of time required for the surgeon to perform
this procedure, $f(\cdot)$ is the target function, and $\eps$ is
noise. We can use a learning procedure to predict $Y$ with $\hat{Y} =
\hat f(X)$ where $\hat f(\cdot)$ is an approximation to
$f(\cdot)$. Given the discussion above, we propose the following
metric for quantifying if this prediction is accurate. We say that the
prediction is accurate (i.e. ``correct'') if
\begin{equation}
  \abs{Y - \hat Y}  < \tau(\hat Y)
\end{equation}
where
\begin{equation}
  \tau(\hat Y) = \min\set{\max\set{p \hat Y, m}, M},
\end{equation}
$p \in (0, 1)$, and $M > m \geq 0$. We see that $\tau(\hat Y)$
encapsulates the issues raised by our hypotheticals: $\tau(\hat Y)$ is
essentially a percentage of $\hat Y$ that is restricted to being
within $[m , M]$. This is depicted in Figure~\ref{fig:thresh}.  

One drawback of this metric is that it is binary and hence treats all
incorrect predictions as equally incorrect. This is an artifact of our
focus on quantifying accuracy rather than error. Given the critical
nature of the patient scheduling, all inaccurate predictions should be
avoided regardless of the the magnitude of inaccuracy.  A benefit of
this approach is that we can easily measure accuracy on the unit
interval. This is particularly helpful when presenting results to
non-technical healthcare professionals.

Another drawback of this performance metric is the induced loss
function:
\begin{equation}
  \begin{split}
\ell(Y, \hat Y) 
= \left\{
  \begin{array}{ll}
    1 &, \abs{Y - \hat Y} \geq \tau(\hat Y)\\
    0 &, \abs{Y - \hat Y} < \tau(\hat Y)
  \end{array}
\right.%
\end{split}
\end{equation}
This loss function is discontinuous and moreover the discontinuity is
sensitive to $p$, $m$, and $M$. This sensitivity is
problematic. Although we developed $\tau(\hat Y)$ (and hence $\ell(Y,
\hat Y)$) based on expert input from healthcare professionals,
translating these qualitative insights into precise parameter values
is fraught with difficulties. Methods like the Delphi technique could
be used to translate expert input into parameter values, but such
methods are not always reliable \cite{Delphi}. 

To alleviate these issues, we would like to ``massage'' this loss
function into a form that is less sensitive to $p$, $m$, and $M$. We
sketch the idea as follows. Suppose that $m$ is sufficiently small and
$M$ is sufficiently large so that $\tau(\hat Y) = p \hat Y$. Then
$\ell(Y, \hat Y) = 0$ when
\begin{equation}
  \abs{Y - \hat Y} < \tau (\hat Y) = p\hat Y.
\end{equation}
Dividing by $\hat Y$ gives us that
\begin{equation}
  \abs{Y/\hat Y - 1} < p
\end{equation}
which is equivalent to
\begin{equation}
  1 - p <Y/\hat Y < 1 + p
\end{equation}
and taking logarithm shows that this is equivalent to
\begin{equation}
  \log(1 - p) < \log(Y) - \log(\hat Y) < \log(1 + p).
\end{equation}
If we let $\eps(p) = \min\set{-\log(1 - p), \log(1 + p)}$, then this
shows that
\begin{equation}
  (\log(Y) - \log(\hat Y))^2 < \eps(p)^2 \implies \ell(Y, \hat Y) = 0.
\end{equation}
So if we aim to minimize $(\log(Y) - \log(\hat Y))^2$ then we will
likely also have $\ell(Y, \hat Y)$ equal to zero\footnote{Technical
  note: One might think that because the logarithm is continuous, we
  can find some $\delta(p) > 0$ such that
  \begin{equation}
    \abs{Y - \hat Y} < \delta(p) \implies \abs{\log(Y) - \log(\hat Y)} < \eps(p)
  \end{equation}
  which would suggest that minimizing $(Y - \hat Y)^2$ will also
  likely give us $\ell(Y, \hat Y) = 0$. However, no such $\delta(p)$
  exists that is independent of both $Y$ and $\hat Y$. This is because
  $x \mapsto \log(x)$ is  not uniformly continuous on
  $(0, \infty)$. See \cite[Chapter~4]{Rudin} for a definition of
  uniform continuity.}.

Although taking the logarithm of $\hat Y$ is not quite the same as
estimating $\log(Y)$, this sketch suggests that given our operational
performance metric, it is reasonable to perform the prediction in
log-space under mean-square loss. Specifically, if we let $Z =
\log(Y)$ then we can use the model
\begin{equation}
  Z = g(X) + \eta
\end{equation}
where $g(\cdot)$ is the target function and $\eta$ is the error. We
can then learn $g(\cdot)$ to get $\hat Z = \hat g(X)$. We can then use
$\exp(\hat Z)$ as a prediction for $Y$.

We note that this sketch merely suggests that using a logarithmic
transformation is appropriate, but it does not provide any guarantees
regarding $\ell(\cdot, \cdot)$. In particular, we see that by doing
the learning in log-space under mean-square loss, we are not taking
into account the asymmetry of the original loss function. Our sketch
suggests that when $m$ is small and $M$ is large this approach is
reasonable but it by no means optimal. Though we sacrifice optimality,
there are several practical benefits to this transformation. As
mentioned earlier, a key benefit is that our models are no longer
sensitive to the parameters $p$, $m$, and $M$ which are necessarily
subjective. Furthermore, by transforming the data and relying on
mean-square loss, we can now apply existing implementations of a
variety of machine learning algorithms. This is useful not only for
research and prototyping but for eventual deployment as well.

\begin{figure}
  \begin{center}
    \subfloat[A histogram of bronchoscopy durations (in minutes). Note the heavy tail and the positive skew.]{\includegraphics[width=\columnwidth]{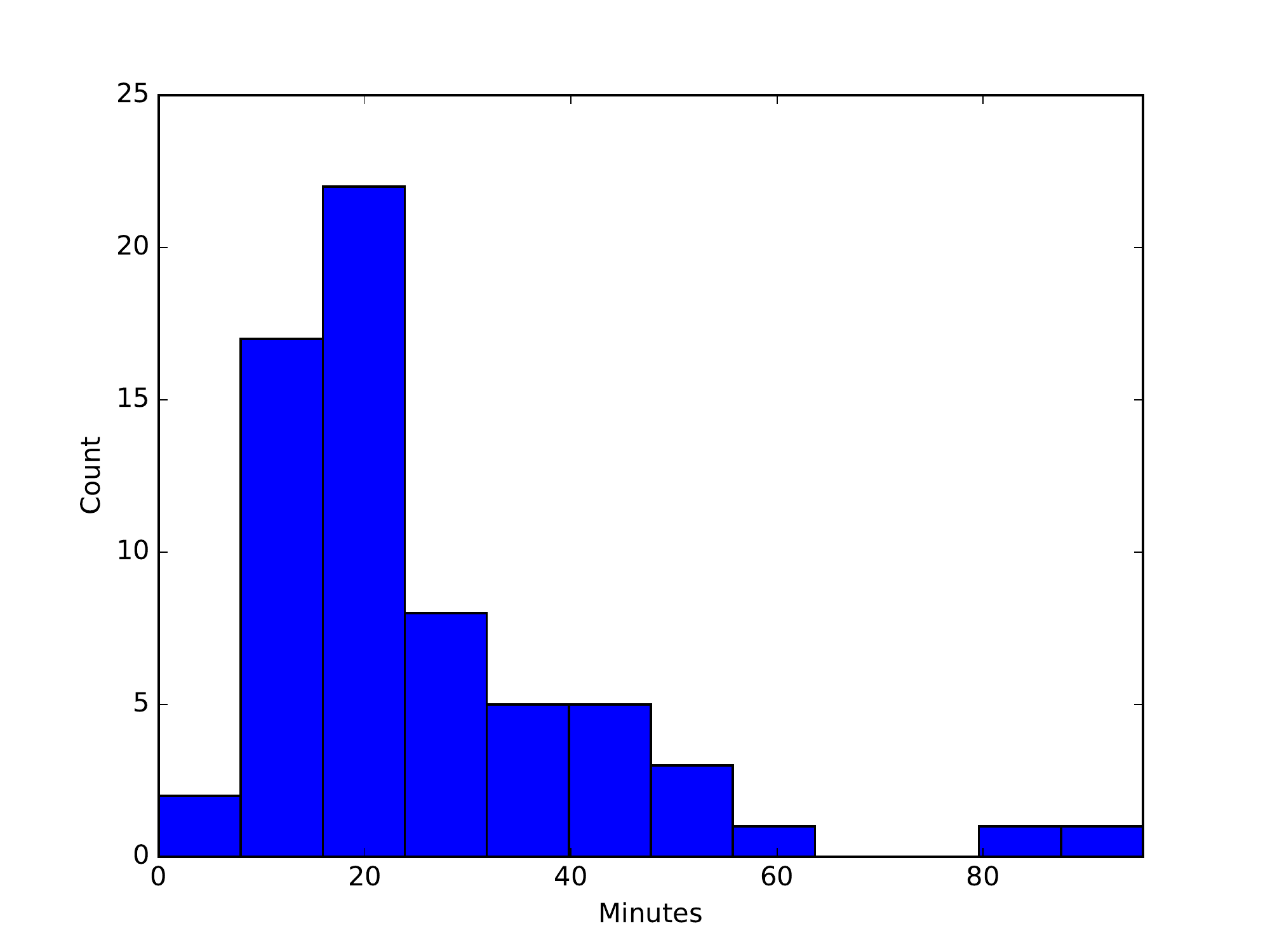}}\\
    \subfloat[A histogram of the natural logarithm of bronchoscopy
    durations. Note that the histogram is fairly
    symmetric.]{\includegraphics[width=\columnwidth]{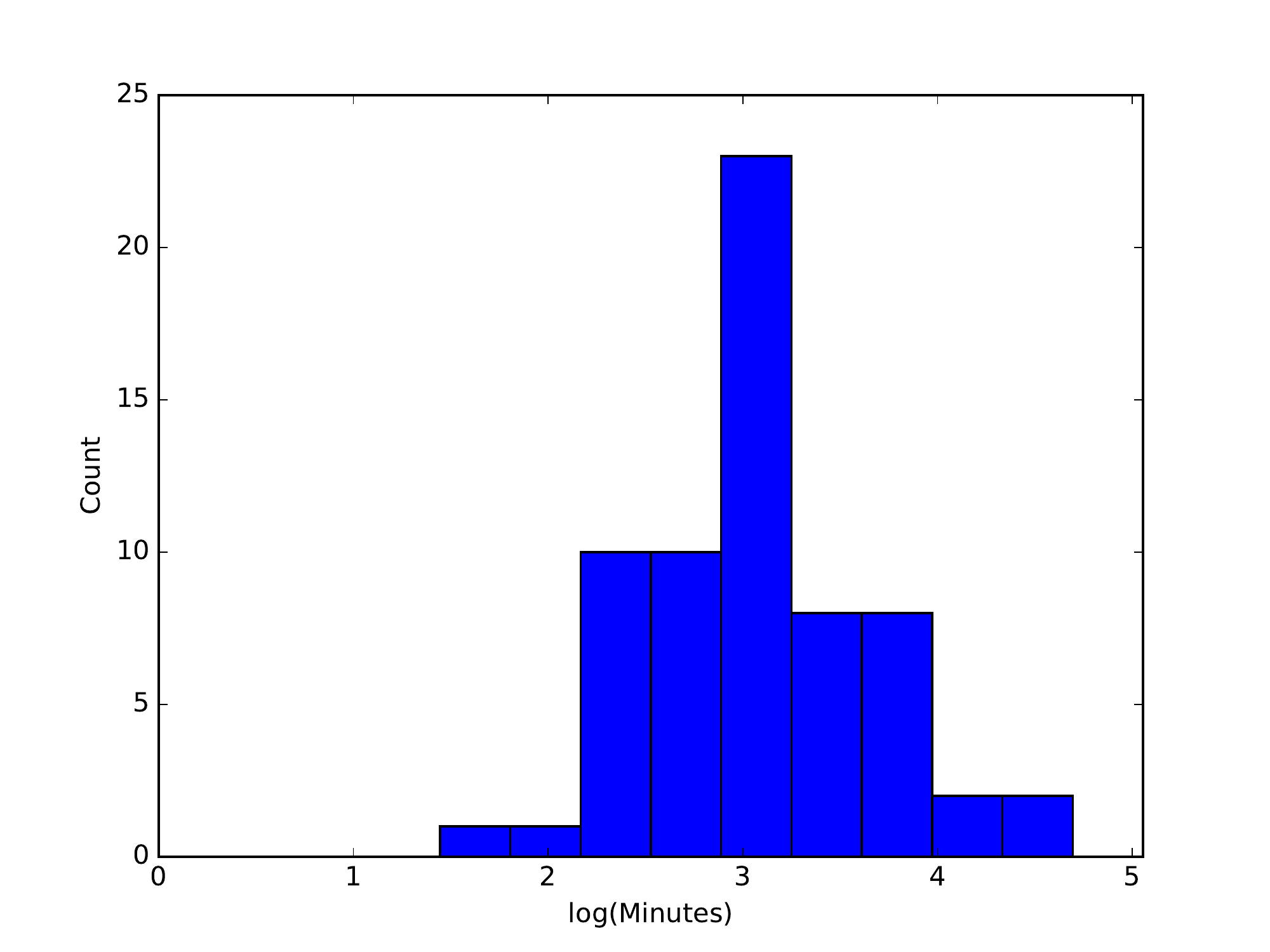}}
  \end{center}
  \caption{A demonstration of the effect of a log transformation on surgical procedure durations.\label{fig:log_density}}
\end{figure}

The logarithmic transformation can also be motivated with more
traditional applied statistics methodology. For example, consider the
histograms in Figure~\ref{fig:log_density}. The original quantity has
a heavy right tail (i.e. a positive skew) but after the logarithmic
tranformation the histogram is fairly symmetric. This also agrees with
the lognormal models discussed in Section~\ref{sec:review}. Although
logarithmic transformations are common, they are not always
acceptable; see \cite{Feng_2013} for some examples of when logarithmic
transformations can actually introduce skew. However, as noted in
\cite{Bland_2013}, for many practical problems logarithmic
transformations can be very useful.

With this discussion in mind, we will train our prediction models in
log-space under mean-square loss to learn $\hat g(\cdot)$. When
evaluating a model on a test set $\set{(Y_i, X_i)}_{i=1}^N$, the
average prediction error will be defined as
\begin{equation}
  \frac{1}{N}\sum_{i=1}^N \ell(Y_i, e^{\hat g(X_i)})
\end{equation}
and the average prediction accuracy will be defined as
\begin{equation}
  \frac{1}{N}\sum_{i=1}^N (1 - \ell(Y_i, e^{\hat g(X_i)})).
\end{equation}
Based on input from nurses and physicians at \LPCH, we have chosen $p =
0.2$, $m = 15$ (minutes), and $M = 60$ (minutes).  We perform a brief
emperical sensitivity analysis in Section~\ref{sec:results}.

\section{Proposed Models and Emperical Results\label{sec:results}}
In this section we describe two benchmark methods that model the
current state of practice (i.e. historical averaging by surgeon and
expert prediction) and propose several tree-based prediction models.
Some of these models provide an automated prediction: they use
features that are available in electronic records and they do not
require input from surgeons. Other models provide a semi-automated
prediction: they use features that are available in electronic records
but they also take advantage of input from surgeons. We use
cross-validation to compare the performance of these models for
several pediatric surgical procedure types. Averaged over
all procedures, the automated ensemble methods are on par with expert
prediction. Given the fact that surgeons have extensive medical
expertise and can potentially use a patient's entire medical record, it
is remarkable that we can achieve this high level of performance
without relying on complicated features. When we augment these
ensemble methods with input from surgeons, we find that semi-automated
prediction significantly outperforms all other methods including
expert prediction.

We note that our models assume the observations are independently and
identically distributed (IID). Although IID assumptions are common, in
most situations such assumptions are idealizations and this situation
is no different. In our case, the IID assumption is reasonable because
patients are treated separately and generally do not impact each other
significantly. However, there can be correlations between their
surgical times if they are scheduled in the same block. For example,
suppose two surgeries are scheduled on the same day for the same
surgeon. If the first surgery runs overtime, the surgeon may feel
pressure to complete the second surgery more quickly. We are assuming
that situations like this are not common and are a second-order
concern.

\subsection{Benchmark Prediction Methods}
As described in Section~\ref{sec:current}, there are currently two
options for predicting surgical case durations:
\begin{enumerate}
\item Historical Averaging: If a surgeon is planning on performing a
  particular surgical procedure, we take the average (arithmetic mean)
  of the last five times the surgeon perfomed this particular
  procedure and use this value as the prediction. In the case that a
  surgeon has not performed this particular surgery at least five
  times, we use the average of all surgeons' past five procedure times
  as the prediction. 
\item Expert Predictions: A surgeon (perhaps with the assistance of
  his staff) gives an expert prediction of how long a surgical
  procedure will last.
\end{enumerate}

Since we are not explicitly modeling any temporal relationship between
observations, historical averaging cannot be directly evaluated and we
instead consider a prediction method \AVG that is used as
follows. Given a training set for a particular procedure, for each
surgeon we record the average amount of time used to complete the
procedure. When predicting how long a procedure in the testing set
will take for a particular surgeon, we use the average computed from
the training set. If there are fewer than five observations for a
particular surgeon in the training set, we average over all surgeons
in the training set.

To model expert predictions, we can use the actual amount of time that
was predicted for each procedure; this amount of time is recorded in
the data set. Since these are the predictions that were used to make
actual scheduling decisions, we refer to this prediction method as
\SCH.

\subsection{Tree-Based Prediction Methods}
Given our critique in Section~\ref{sec:review} of the regression tree
method used in \cite{Bravo_2015}, we propose three tree-based
automated prediction methods. Motivated by the discussion in
Section~\ref{sec:metric}, for each of these methods, we perform the
prediction in log-space and transform the result back to a linear
scale by exponentiating.

\begin{table}
  \begin{center}
    \begin{tabular}{|C{0.18\columnwidth}|C{0.33\columnwidth}|C{0.33\columnwidth}|}
      \hline
      ASA Classification & Definition & Examples\\       
      \hhline{|=|=|=|}
      ASA I & A normal healthy patient & Healthy, non-smoking, no/minimal alcohol use\\
      \hline
      ASA II & A patient with mild systemic disease & Smoking, social alcohol drinker, obesity\\
      \hline
      ASA III & A patient with severe systemic disease & Active hepatitis, alcohol dependence/abuse, morbid obesity\\
      \hline
      ASA IV & A patient with severe systemic disease that is a constant threat to life & Ongoing cardiac ischemia or severe valve dysfunction, sepsis\\
      \hline
      ASA V & A moribund patient who is not expected to survive without the operation & A ruptured abdominal/thoracic aneurysm, massive trauma, intracranial bleed with mass effect\\
      \hline
      ASA VI & A declared brain-dead patient whose organs are being removed for donor purposes & {}\\
      \hline
    \end{tabular}
  \end{center}
  \caption{Definition of the American Society of Anesthesiologists (ASA)
    Physical Status Classification System as described in \cite{ASA_Online}\label{tab:ASA}}
\end{table}

Each of the proposed models uses the following features:
\begin{itemize}
\item Gender of the patient (male vs. female)
\item Weight of the patient (in kilograms)
\item Age of the patient (in years)
\item American Society of Anesthesiologists (ASA) physical
  status/score of the patient as described in
  Table~\ref{tab:ASA}\footnote{In our data set, there were no patients
    with ASA VI.}
\item Primary surgeon identity
\item Location (in an OR vs. in the APU)
\item Patient class (in-patient vs. out-patient)
\item Procedure name
\end{itemize}
We did not ``mine'' our data to choose these features; each of these
features is motivated by our domain knowledge.  For example, the first
four features (gender, weight, age, and ASA score) provide a crude
summary of the patient's clinical state. Although \cite{Bravo_2015}
reported that the surgeon identity was not useful, conventional wisdom
suggests that surgeon identity is useful and so we opt to include it
as a feature. The location and patient classification provide some
basic information about the expected complexity of the procedure --
procedures performed in the APU are typically shorter and simpler;
out-patient procedures also tend to be less complex.

The procedure name has obvious predictive power but is actually quite
nuanced. The procedure display names that are currently used for
operational purposes do not necessarily fully distinguish different
procedures. For example, 18 cases in our data set were scheduled with
the procedure name ``Radiation Treatment.''  This name does not include
the type of radiation treatment (i.e. internal vs. external) or the
part of the body. Current Procedural Terminology (CPT) codes provide a
detailed and standardized way of describing procedures. Although a set
of potential CPT codes is known to the surgeon \textit{ex~ante}, the
particular CPT code used is only recorded
\textit{ex~poste}. Consequently, we rely on the procedure display name
rather than CPT code as a feature.

Each of the proposed models is based on regression trees. The simplest
model is a single decision tree regressor \cite{CART}, denoted
\DTR. We also consider ensembles of trees. In particular, we use a
random forest regressor \cite{Breiman_RF}, denoted \RFR, and an
ensemble of adaptively boosted regression trees \cite{Drucker_1997,
  Elith_2008}, denoted \ABR. For each of these methods, we rely on the
implementations provided by the \texttt{scikit-learn} package
\cite{Scikit}\footnote{We mostly use the default parameters but we set
  \texttt{min\_samples\_split} equal to 10. This is a simple way of
  preventing a single decision tree from overfitting; this parameter
  choice does not significantly affect ensembles of decision
  trees.}. Note that while \DTR may seem like the same method that was
used (unsuccessfully) in \cite{Bravo_2015}, recall that we are fitting
our trees in log-space and we are also measuring performance according
to an alternative criterion.

\DTR, \RFR, and \ABR each provide an automated prediction method: the
aforementioned features are easily pulled from electronic medical
records and can be plugged into the learned models. However, we can
also use these methods in a semi-automated fashion. In addition to the
aforementioned features, we can also use the prediction provided by
the surgeon as a feature. The idea is that the surgeon can still
provide expert input and the model can use the other features to
adjust the expert prediction.  Since the expert prediction is the
output of \SCH, we refer to \DTRSCH, \RFRSCH, and \ABRSCH as \DTR,
\RFR, and \ABR with the additional feature of the expert
prediction. The potential benefit of this approach is improved
prediction accuracy, but we immediately lose the benefits of
automation. Another downside of this approach is concept drift:
surgeon behavior may adapt and the model may need to be perodically
re-trained.  We discuss these issues more in Section~\ref{sec:future}.

\subsection{Prediction Results}

\begin{table*}
  \begin{center}
    \resizebox{\textwidth}{!}{%
      \begin{tabular}{|C{0.19\textwidth}||c|c|c|c|c|c|c|c|c|c|c|}
        \hline
        {} & $N$ & $\hat\mu$ & $\hat\sigma$ & \AVG & \SCH & \DTR & \RFR & \ABR & \DTR-\SCH & \RFR-\SCH & \ABR-\SCH\\
        \hhline{|=#=|=|=|=|=|=|=|=|=|=|=|}

        Overall & 917 & 44.20 & 36.56 & 
        0.44 (0.08) & 0.62 (0.04) & 0.59 (0.03) & 0.65 (0.04) & 0.65 (0.04) & 0.64 (0.04) & 0.71 (0.04) & 0.70 (0.03)\\
        \hhline{|=#=|=|=|=|=|=|=|=|=|=|=|}

        Adenoidectomy & 56 & 29.36 & 25.86 & 
        0.53 (0.18) & 0.73 (0.13) & 0.60 (0.15) & 0.70 (0.15) & 0.71 (0.13) & 0.67 (0.16) & 0.77 (0.15) & 0.76 (0.14) \\ 
        \hline

        Bilateral Ear Myringotomy with Tubes & 43 & 28.60 & 34.30 & 
        0.30 (0.20) & 0.74 (0.10) & 0.66 (0.13) & 0.71 (0.16) & 0.72 (0.15) & 0.74 (0.17) & 0.82 (0.12) & 0.80 (0.11) \\ 
        \hline

        Bone Marrow Aspiration & 42 & 39.83 & 28.11 & 
        0.55 (0.18) & 0.65 (0.16) & 0.50 (0.18) & 0.59 (0.20) & 0.63 (0.17) & 0.54 (0.16) & 0.60 (0.18) & 0.65 (0.16) \\ 
        \hline

        Bronchoscopy (Pulmonary) & 55 & 33.33 & 21.33 & 
        0.48 (0.21) & 0.42 (0.11) & 0.60 (0.19) & 0.66 (0.13) & 0.64 (0.15) & 0.53 (0.14) & 0.68 (0.13) & 0.66 (0.14) \\ 
        \hline

        Colonoscopy with Biopsy & 50 & 71.32 & 25.71 & 
        0.36 (0.13) & 0.42 (0.12) & 0.39 (0.15) & 0.45 (0.16) & 0.46 (0.18) & 0.37 (0.16) & 0.47 (0.15) & 0.43 (0.16) \\ 
        \hline

        Dental Rehabilitation & 78 & 121.03 & 41.72 & 
        0.06 (0.10) & 0.50 (0.11) & 0.41 (0.12) & 0.45 (0.14) & 0.46 (0.13) & 0.43 (0.11) & 0.48 (0.11) & 0.50 (0.11) \\ 
        \hline

        Esophagogastroduodenoscopy (EGD) with Biopsy & 105 & 34.94 & 23.97 & 
        0.37 (0.14) & 0.63 (0.11) & 0.49 (0.11) & 0.58 (0.10) & 0.55 (0.08) & 0.73 (0.12) & 0.74 (0.11) & 0.75 (0.11) \\ 
        \hline

        Laparoscopic Appendectomy & 81 & 54.99 & 16.52 & 
        0.42 (0.16) & 0.27 (0.10) & 0.58 (0.14) & 0.63 (0.14) & 0.63 (0.13) & 0.54 (0.14) & 0.64 (0.13) & 0.62 (0.12) \\ 
        \hline

        Lumbar Puncture with Intrathecal Chemotherapy & 50 & 39.54 & 31.98 & 
        0.30 (0.14) & 0.63 (0.15) & 0.47 (0.17) & 0.43 (0.17) & 0.52 (0.18) & 0.62 (0.15) & 0.69 (0.16) & 0.74 (0.13) \\ 
        \hline

        Myringotomy with Tubes & 86 & 32.74 & 37.73 & 
        0.30 (0.18) & 0.73 (0.12) & 0.63 (0.12) & 0.66 (0.10) & 0.65 (0.12) & 0.73 (0.11) & 0.76 (0.09) & 0.73 (0.11) \\ 
        \hline

        Portacath Removal & 56 & 31.98 & 12.20 & 
        0.76 (0.15) & 0.93 (0.08) & 0.70 (0.15) & 0.78 (0.13) & 0.78 (0.11) & 0.71 (0.14) & 0.77 (0.15) & 0.76 (0.13) \\ 
        \hline

        Tonsillectomy And Adenoidectomy & 215 & 29.96 & 13.05 & 
        0.62 (0.15) & 0.71 (0.07) & 0.73 (0.05) & 0.78 (0.05) & 0.77 (0.07) & 0.74 (0.07) & 0.79 (0.05) & 0.78 (0.05) \\ 
        \hline

      \end{tabular}
    }
  \end{center}
  \caption{Estimated Average Prediction Accuracy. For each procedure we report the sample size ($N$), the emperical mean ($\hat\mu$), the emperical standard deviation $(\hat\sigma)$, and the estimated average prediction accuracy of each method. A prediction is declared accurate based on the metric discussed in Section~\ref{sec:metric}. We use $5 \times 5$-fold cross-validation to estimate the mean prediction accuracy and we also report the standard error of this estimate.\label{tab:results}}
\end{table*}

Our data set includes all surgical procedures performed at \LPCH from
May 2014 through January 2016. This data set includes 4475 unique
procedure names, but we need to pare this list to avoid the small
sample problems in \cite{Bravo_2015}. We only consider procedures for
which we have at least 40 observations. There are only 12 procedure
names that meet this restriction; these procedures are listed
alphabetically in Table~\ref{tab:results}. Overall, we have 917
observations which is sufficient for fitting our tree-based
models. Although this sample size restriction limits the breadth of
our study, it also focuses our study on procedure types with the most
significant operational impact. In addition, we note that our chosen
features are not specific to these procedure types and so our
conclusions should generalize to other procedures.

We use $5 \times 5$-fold cross-validation to estimate the average
prediction accuracy for each method and also provide a breakdown based
on each procedure name; the results are shown in
Table~\ref{tab:results}. We use the shorthand ``Method 1 $\prec$
Method 2'' to indicate that the estimated mean prediction accuracy of
Method 1 is less than or equal to Method 2. We also describe this as
Method 2 outperforming Method 1.  Overall, we see that
\begin{align*}
  &\AVG \prec \DTR \prec \SCH \\
  &\prec \DTRSCH \prec \RFR \prec \ABR \prec \ABRSCH \prec \RFRSCH
\end{align*}
with \RFR and \ABR achieving the same estimated mean prediction
accuracy. Although \DTR does not outperform expert prediction, \RFR
and \ABR are able to outperform expert predictions, albeit only
slightly. By including expert predictions as a feature to these
methods, we significantly increase prediction accuracy with the
semi-automated prediction models \DTRSCH, \RFRSCH, and \ABRSCH all
outperforming their automated counterparts. By including expert
information, \RFRSCH and \ABRSCH both outperform \SCH significantly.

We gain additional insights by breaking down the results by procedure;
we first compare the automated methods to the benchmarks. We see that
\DTR outperforms \AVG for 10 of the procedures. This shows that while
\DTR is better than \AVG overall, it is not better for all
procedures. We also see that \SCH outperforms \AVG for only 10
procedures so even expert prediction is not always better than
historical averaging. In contrast, \RFR and \ABR outperform \AVG for
all procedures. This shows that either \RFR or \ABR could be used as
replacements for historical averaging as an automated prediction
method.

Although \RFR and \ABR outperform \AVG, it is less clear whether they
truly outperform \SCH.  Although \RFR and \ABR slightly outperform
\SCH overall, \RFR and \ABR each outperform \SCH for only 4
procedures. When \RFR and \ABR outperform \SCH, they often do so by a
large margin. For example, when predicting bronchoscopy durations,
\RFR and \ABR outperform \SCH by a factor of 3. In contrast, when \SCH
outperforms \RFR and \ABR, the margin is often small. For example,
when predicting adenoidectomies, \SCH only barely outperforms \RFR and
\ABR. However, \SCH can sometimes outperform \RFR and \ABR by a large
margin; for example, this is the case for portacath removals. Because
the results are somewhat mixed, it seems that \RFR and \ABR should not
replace expert predictions outright; if we wanted to rely on an
automated prediction method, it would be more appropriate for \RFR or
\ABR to replace expert predictions only for certain procedures. 

Now consider the semi-automated prediction methods, \DTRSCH, \RFRSCH,
and \ABRSCH. All of these methods outperform \AVG for each procedure
and overall. They outperform \SCH overall, but again we need to
breakdown the results procedure by procedure. \DTRSCH outperforms \SCH
for only 6 procedures so \DTRSCH and \SCH are best viewed as
comparable. However, \RFRSCH outperforms \SCH for 9 procedures and
\ABRSCH outperforms \SCH for 11 procedures. This shows that the
semi-automated ensemble methods offer an improvement in prediction
accuracy over raw expert predictions. Of course, \RFRSCH and \ABRSCH
are potentially more expensive than \RFR and \ABR (they require input
from surgeons), but depending on hospital needs the improvement in
prediction accuracy may be worth it.

\subsection{Feature Importance}
\begin{table*}
  \begin{center}
    \begin{tabular}{|c||c|c|c||c|c|c|}
      \hline
      {} & \DTR & \RFR & \ABR & \DTRSCH & \RFRSCH & \ABRSCH\\
      \hhline{|=#=|=|=#=|=|=|}
      Gender            & 0.01 & 0.01 & 0.02 & 0.01 & 0.01 & 0.02\\
      \hline
      Weight            & 0.15 & 0.15 & 0.25 & 0.08 & 0.08 & 0.19\\
      \hline
      Age               & 0.08 & 0.08 & 0.11 & 0.05 & 0.05 & 0.09\\
      \hline
      ASA Score         & 0.03 & 0.04 & 0.05 & 0.02 & 0.02 & 0.03\\
      \hline
      Primary Surgeon   & 0.16 & 0.19 & 0.24 & 0.07 & 0.07 & 0.11\\
      \hline
      Location          & 0.10 & 0.08 & 0.06 & 0.01 & 0.01 & 0.03\\
      \hline
      Patient Class     & 0.04 & 0.05 & 0.05 & 0.02 & 0.03 & 0.04\\
      \hline
      Procedure Name    & 0.42 & 0.41 & 0.22 & 0.12 & 0.10 & 0.10\\
      \hline
      Expert Prediction & N/A  & N/A  & N/A  & 0.62 & 0.62 & 0.38\\
      \hline
    \end{tabular}
  \end{center}
  \caption{Average Relative Feature Importance~\label{tab:importance}}
\end{table*}

Because we are using tree-based methods, we can also use the mean
decrease in risk across splits as a heuristic for relative feature
importance \cite{CART}. For each method, this heuristic provides a
non-negative score for each feature with these scores summing to
one. We average the relative importance across the $5\times5$
cross-validation and show the results in
Table~\ref{tab:importance}. Note that because the semi-automated
methods have an additional feature, the relative importance scores of
the automated methods should not be compared directly to the relative
importance scores of the semi-automated methods.

First consider the automated methods. For \DTR, \RFR, and \ABR, the
procedure name, patient weight, and primary surgeon identity are the
most important features. Procedure name and primary surgeon identity
are basic pieces of information that have obvious predictive value;
indeed, this is why historical averaging is currently so common. This
contradicts the conclusion in \cite{Bravo_2015} that surgeon identity
is not a useful feature.

\begin{figure}
  \begin{center}
    \includegraphics[width=\columnwidth]{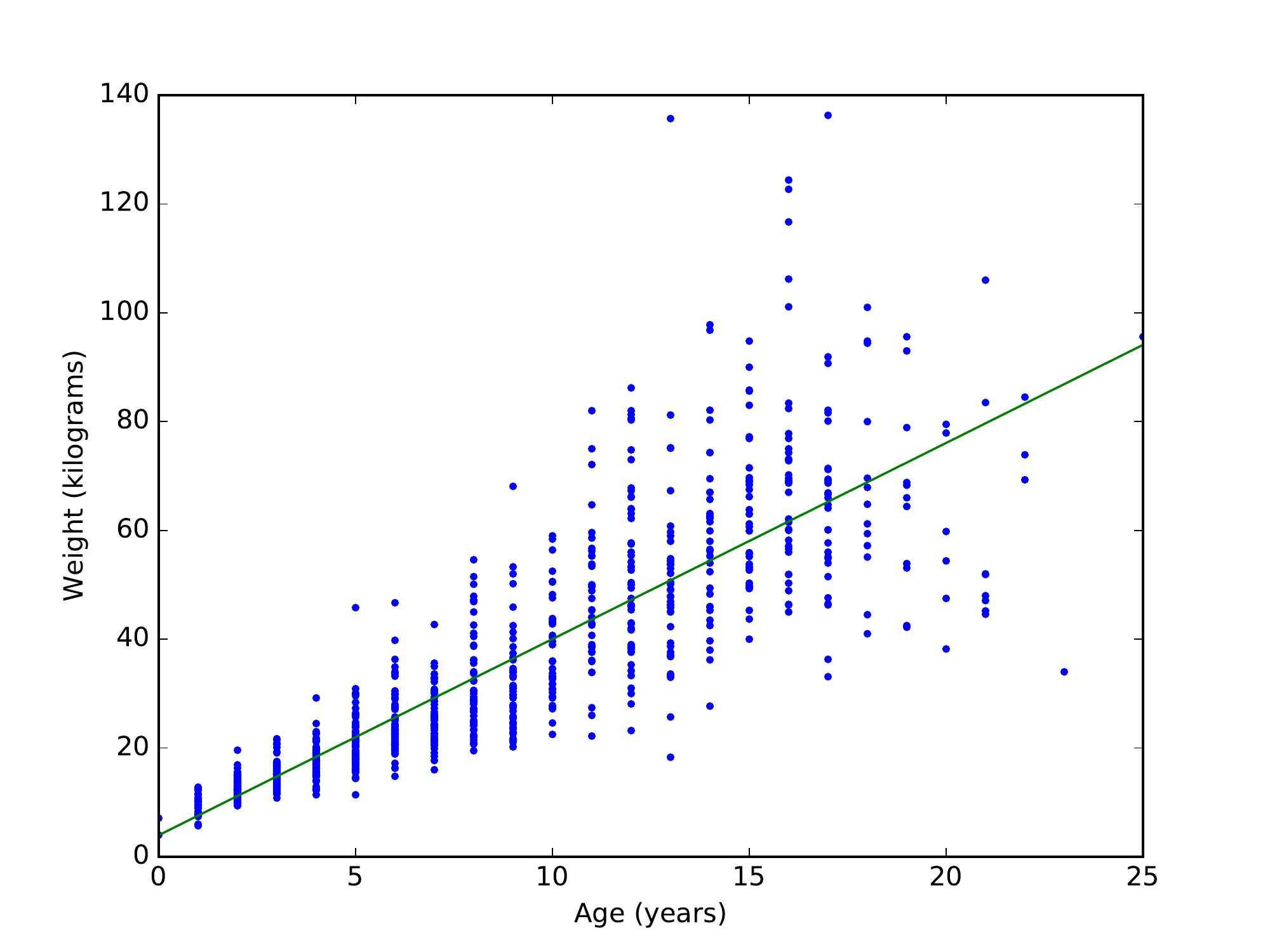}
  \end{center}
  \caption{Weight vs. Age. A scatter plot along with the ordinary
    least squares regression line. \label{fig:weight_v_age}}
\end{figure}

It may be surprising that patient weight is such an important feature,
but we offer two explanations. We first note that age is typically
used to indicate developmental status in children and weight
correlates strongly with age; in our data the Pearson correlation
coefficient between weight and age is 0.85. This also explains why age
does not have as high an importance score. Secondly, we note that
childhood obesity is an increasingly pervasive public-health issue
\cite{Ebbeling_2002} and obesity is known to lead to complications
during surgery \cite{Byrne_2001}. These ideas are supported by
Figure~\ref{fig:weight_v_age} which shows patient weight as a function
of age. Figure~\ref{fig:weight_v_age} shows a strong correlation
between age and weight but it also shows that the distribution of
weights is positively skewed, particularly for teenage patients.

We can also gain insights about the features with low relative
importance. Recall that location and patient class are included as
features because they contain some information about the complexity of
the operation. Although each of these features has fairly low
importance, for \RFR and \DTR the combined importance of these
features is comparable to the importance of patient weight. This
suggests that location and patient class are fairly effective proxies
for procedure complexity. We see that the patient ASA score has a low
relative importance. We conjecture that this is because information
encoded in the ASA score is better represented by other features. In
particular, Table~\ref{tab:ASA} shows that obesity is part of the ASA
score, but this information is better represented by the patient
weight.  We also see that patient gender has low predictive
power. Patient gender is typically not a useful predictor for surgical
times; in fact, patient gender was not used in \cite{Bravo_2015}.

Now consider the semi-automated methods. We see that expert prediction
is by far the most important feature to \DTRSCH, \RFRSCH, and
\ABRSCH. However, the next three most important features are procedure
name, primary surgeon, and weight. We also see the same trend that
gender and ASA score are not very important features. 

\subsection{Sensitivity to the Performance Metric}
\begin{figure}
  \begin{center}
    \includegraphics[width=\columnwidth]{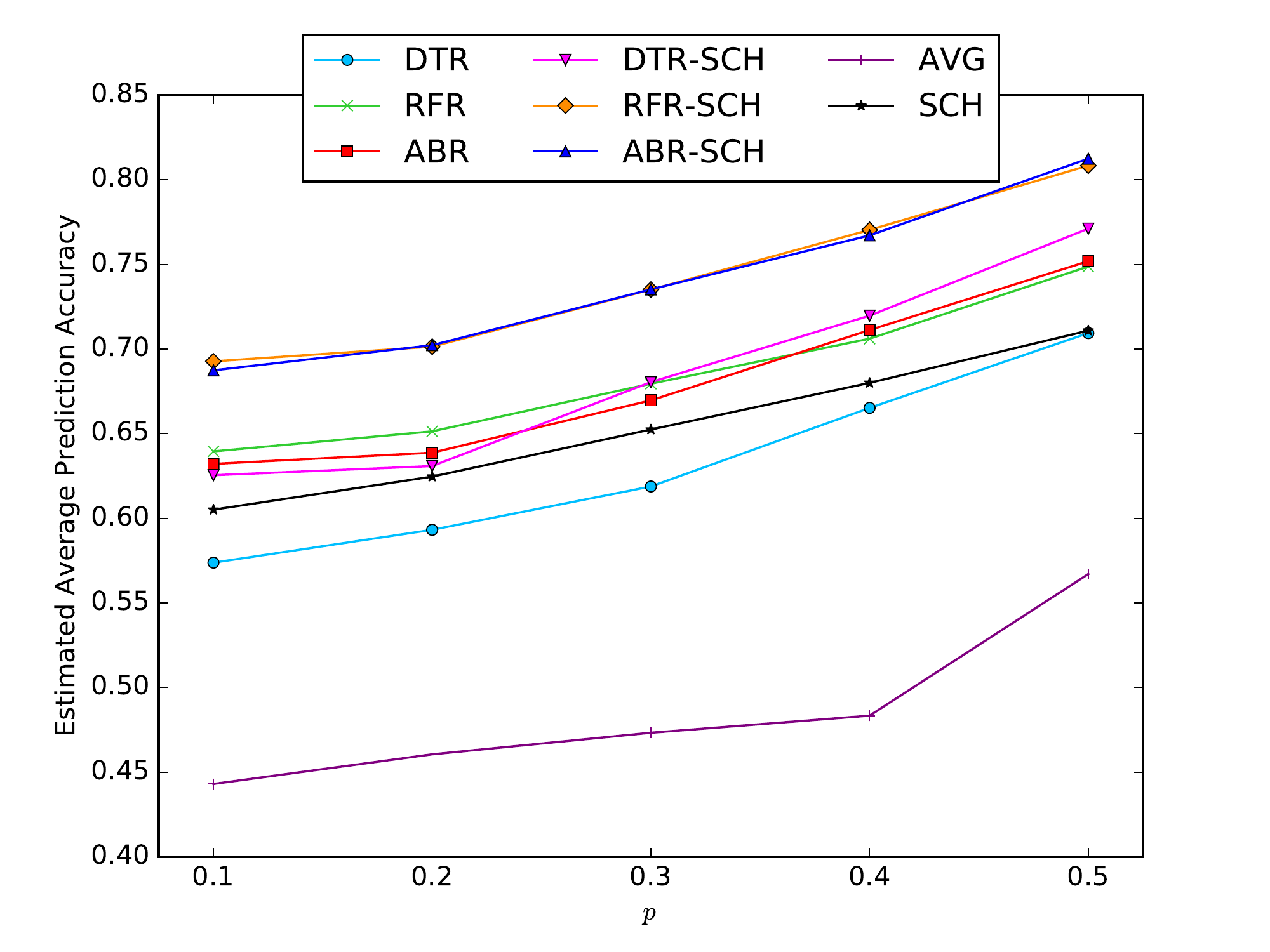}
  \end{center}
  \caption{Estimated Overall Prediction Accuracy vs. $p$. We vary $p$
    while holding $m = 15$ and $M = 60$.\label{fig:varyp}}
\end{figure}

Finally, we make a brief comment regarding the performance metric. As
noted in Section~\ref{sec:metric}, the choice of $p$, $m$, and $M$ is
inherently subjective and the estimated prediction accuracy of each
method is sensitive to these parameters. In Figure~\ref{fig:varyp}, we
plot the estimated prediction accuracy as $p$ varies with $m = 15$ and
$M = 60$ fixed. As $p$ increases, the performance requirements become
more lax and the estimated accuracies of all methods generally
increase. We note the following trends:
\begin{itemize}
\item \AVG is by far the least accurate method
\item \DTR is much better than \AVG
\item \SCH is slightly better than \DTR
\item \RFR, \ABR, and \DTRSCH outperform \SCH and have comparable
  performance for all $p$
\item \RFRSCH and \ABRSCH are by far the most accurate methods and
  have comparable performance for all $p$
\end{itemize}
This suggests that although the estimated prediction accuracy depends
on the choice of parameters, the general trends that we've noted
should hold for a wide range of parameter choices.

\section{Directions of Future Work\label{sec:future}}
Our current work suggests some new directions.  Although we chose to
focus on tree-based methods to draw a contrast with the negative
results from \cite{Bravo_2015}, there are other nonparametric
regression methods (e.g. nearest neighbor regression and kernel
regression \cite{Altman_1992}) that are also worth exploring. In
addition, given that different methods will have higher predictive
accuracy for different procedures, it may be worth determining which
methods are best suited for different scenarios. 

There are also many other methods of incorporating expert opinions
into prediction models. In particular, Bayesian methodologies could
provide a rigorous framework for incorporating expert knowledge from
surgeons and nurses. Bayesian methods can also help us deal with
smaller samples sizes. This can help us broaden the applicability of
our results to procedures that are less common.

As noted above, when incorporating expert opinions we increase the
possibility of concept drift: surgeons may adapt to the semi-automated
methods so that the prediction accuracy degrades. There are different
ways of handling concept drift, e.g. \cite{Klinkenberg_2000,
  Kolter_2005}, which should be explored as we work towards a
deployment.

In addition to considering different methods, it may also be useful to
consider different features. Our current feature set is intentionally
generic: the features that we consider can be used at any pediatric
hospital and for any procedure, giving our models broad applicability.
However, it may be worth sacrificing this broad applicability to use
more specific features that yield improved prediction accuracy. For
example, in teaching hospitals it is known that having a resident in
the OR will lead to longer surgeries \cite{Vinden_2016,
  Bridges_1999}. For specific procedures, it may be useful to have
more detailed clinical information about the patient. Feature
engineering can be an open-ended process but more extensive feature
engineering is likely to improve the predictive power of our models.

\section{Conclusions\label{sec:conclusions}}
Motivated by operational problems in hospitals, we have studied the
problem of building prediction models for pediatric surgical case
durations. We have proposed a novel performance metric for prediction
in this application. Not only does this performance metric capture
issues relevant to hospital operations, it also motivates a nonlinear
transformation of the data. In light of the negative results in the
medical literature, we opt to focus on tree-based prediction
models. We demonstrate that contrary to the medical literature, our
models outperform currently used algorithms and are often on par with
human experts. When we take advantage of expert opinions, our models
can significantly outperform surgeons. These positive results point to
new directions of research that will ultimately enable automated and
semi-automated prediction methods to be deployed in pediatric
hospitals.

\ACKNOWLEDGEMENTS

\bibliographystyle{ieeetr}
\bibliography{Master_DSAA2016}

\end{document}